\documentclass[prl,twocolumn,showpacs,superscriptaddress,floatfix]{revtex4}
\usepackage{graphicx,amssymb,amsmath}

\newlength{\ldag}
\newcommand{\cra}{a^\dagger}
\newcommand{\ana}{{a^{\phantom\dagger}\hspace{-\ldag}}}
\newcommand{\crb}{b^\dagger}
\newcommand{\anb}{{b^{\phantom\dagger}\hspace{-\ldag}}}

\newcommand{\comment}[1]{}

\begin{document}
\title{Entanglement entropy beyond the free case}

\author{Thomas Barthel}
\affiliation{Institute for Theoretical Physics C, RWTH Aachen, 52056 Aachen, Germany}

\author{S\'ebastien Dusuel}
\affiliation{Lyc\'ee Louis Thuillier, 70 Boulevard de Saint Quentin, 80098 Amiens Cedex 3, France}

\author{Julien Vidal}
\affiliation{Laboratoire de Physique Th\'eorique de la Mati\`ere Condens\'ee, CNRS UMR 7600,
Universit\'e Pierre et Marie Curie, 4 Place Jussieu, 75252 Paris Cedex 05, France}

\begin{abstract}
We present a perturbative method to compute the ground state entanglement entropy for interacting systems. We apply it to a collective model of mutually interacting spins in a magnetic field. At the quantum critical point, the entanglement entropy scales logarithmically with the subsystem size, the system size, and the anisotropy parameter. We determine the corresponding scaling prefactors and evaluate the leading finite-size correction to the entropy. Our analytical predictions are in perfect agreement with numerical results.
\end{abstract}
%
%

%\pacs{21.60.Fw, 21.10.Re, 75.40.Cx, 73.43.Nq, 05.10.Cc}
\maketitle
%
%
%%%%%%%%%%%%%%%%%%%%%%%%%%%%%%%
%%%%%%%%%%%%%%%%%%%%%%%%%%%%%%%
%  Introduction
%%%%%%%%%%%%%%%%%%%%%%%%%%%%%%%
%%%%%%%%%%%%%%%%%%%%%%%%%%%%%%%
%
%
In recent years, much effort has been devoted to the characterization of ground state entanglement in many-particle systems. Especially, its relationship with quantum phase transitions (QPTs) has been investigated, following the seminal works in one-dimensional (1D) systems \cite{Osborne02,Osterloh02,Latorre03}. Entanglement is also a key concept for quantum information theory \cite{Nielsen00}.
Several fundamental questions have emerged concerning the universality of the observed behaviors, as well as their classification.  
For example, the entanglement entropy is known to scale logarithmically with the subsystem size in 1D critical spin chains \cite{Holzhey94,Latorre03,Korepin04,Refael04,Eisert06}, though its precise form depends on the boundary conditions \cite{Calabrese04,Zhou06}.
It is natural to wonder how that behavior is modified in higher-dimensional systems.

To address this question, we consider the Lipkin-Meshkov-Glick model (LMG) \cite{Lipkin65} of mutually interacting spins in a magnetic field, i.e.\ a system with an infinite coordination number.
Although the model was introduced in nuclear physics, it has been used to describe many other physical systems such as Bose-Einstein condensates \cite{Cirac98} or small ferromagnetic particles \cite{Chudnovsky88} to cite just a few.
Its entanglement properties have been analyzed from different perspectives \cite{Vidal04_1,Dusuel04_3,Dusuel05_2}, but its entropy has only been studied numerically \cite{Latorre05_2}. 

The aim of this Letter is to investigate analytically the entanglement entropy in the LMG model.
First, the model is studied in the thermodynamic limit for which it can be mapped onto a free 
bosonic system, allowing for an exact evaluation of the entropy \cite{Peschel99,Peschel03,Barthel06_1}.
In a second step, we address the finite-size corrections to the entropy at and away from the transition point. This leads us to introduce a perturbative method, because for a large but finite number of spins, the bosons are weakly interacting.
At the critical point, the entropy is found to grow logarithmically with the 
subsystem size as in the critical 1D $XY$ model \cite{Peschel04,Its05} which is the 1D counterpart of the LMG model.
However, the scaling prefactor differs from the 1D case and also disagrees 
with previous numerical studies \cite{Latorre05_2}. We also show that, at the 
quantum critical point, the entropy scales logarithmically with the system 
size and the anisotropy parameter, and we compute the associated scaling 
prefactors. Away from criticality, the dependence of the entropy on the subsystem size is found to differ considerably from the scaling in finite-dimensional systems.
The validity of our approach is confirmed by numerical exact diagonalization results.   

%
%
%%%%%%%%%%%%%%%%%%%%%%%
%%%%%%%%%%%%%%%%%%%%%%%
\emph{The model ---}
%%%%%%%%%%%%%%%%%%%%%%%
%%%%%%%%%%%%%%%%%%%%%%%
%
The LMG model describes the collective behavior of $N$ spins $1/2$ with a mutual anisotropic ($XY$) ferromagnetic interaction, subjected to a transverse magnetic field $h$.
Introducing the total spin operators $S_{\alpha}=\sum_{i} \sigma_{\alpha}^{i}/2$, where $\sigma_{\alpha}$ are the Pauli matrices, and the anisotropy parameter $\gamma$, the Hamiltonian  reads
%
%
%%%%%%%%%%
\begin{equation}
  H=-\frac{1}{N} \big(S_x^2 + \gamma S_y^2\big) - h \: S_z.
    \label{eq:hamiltonian}
\end{equation}
%%%%%%%%%%
%
% 
Without loss of generality, we can assume $0 \leqslant \gamma < 1$ and $h \geqslant 0$.
As discussed in the literature, this system undergoes a second-order QPT at $h=1$, between a symmetric ($h>1$) and a broken ($h<1$) phase, which is well described by a mean-field approach. The corresponding classical ground state is fully polarized in the field direction ($\langle\sigma_z^i\rangle$=1) for $h>1$, and twofold degenerate with $\langle\sigma_z^i\rangle$=$h$ for $h<1$.
Entanglement originates from quantum fluctuations around these classical ground states.

%
%
%%%%%%%%%%%%%%%%%%%%%%%
%%%%%%%%%%%%%%%%%%%%%%%
\emph{The entanglement entropy ---}
%%%%%%%%%%%%%%%%%%%%%%%
%%%%%%%%%%%%%%%%%%%%%%%
%
We wish to analyze the entanglement entropy of the ground state $|\psi\rangle$, defined as
%
%
%%%%%%%%%%
\begin{equation}
  \mathcal{E}=-\mathrm{Tr}_\mathcal{A}\left(\rho_\mathcal{A}\ln\rho_\mathcal{A}\right)
            =-\mathrm{Tr}_\mathcal{B}\left(\rho_\mathcal{B}\ln\rho_\mathcal{B}\right),
    \label{eq:def_entropy}
\end{equation}
%%%%%%%%%%
%
% 
where $\rho_{\mathcal{A},\mathcal{B}}=\mathrm{Tr}_{\mathcal{B},\mathcal{A}} \left(|\psi\rangle\langle\psi|\right)$. 
This definition relies on a splitting of the system into two blocks $\mathcal{A}$ and $\mathcal{B}$, of sizes $L$ and $(N-L)$ respectively. Here, it amounts to introduce $S_{\alpha}^{\mathcal{A},\mathcal{B}}=\sum_{i \in {\mathcal{A},\mathcal{B}}} \sigma_{\alpha}^{i}/2$. 
To describe quantum fluctuations and thus to compute the entanglement entropy, it is convenient to use the Holstein-Primakoff representation of these spin operators \cite{Holstein40}
%
%
%%%%%%%%%%%%%%%%%%%%
       \begin{eqnarray}
         S_z^\mathcal{A}&=&L/2 - \cra \ana , \label{eq:HP1A} \\
         S_-^\mathcal{A}&=&\sqrt{L}  \: \cra  \: \sqrt{1- \cra \ana/L}=(S_+^\mathcal{A})^\dag, 
         \label{eq:HP2A} 
         \end{eqnarray}
         \begin{eqnarray}
         S_z^\mathcal{B}&=&(N-L)/2 - \crb \anb , \label{eq:HP1B} \\
         S_-^\mathcal{B}&=&\sqrt{N-L} \: \crb \: \sqrt{1-\crb \anb/ (N-L)}=(S_+^\mathcal{B})^\dag, 
         \label{eq:HP2B}
       \end{eqnarray}
%%%%%%%%%%%%%%%%%%%%
%
%
with $S_\pm^{\mathcal{A},\mathcal{B}}=S_x^{\mathcal{A},\mathcal{B}} \pm {\rm i}\,  S_y^{\mathcal{A},\mathcal{B}}$. In this way, the LMG Hamiltonian is mapped onto a system of two interacting bosonic modes $a$ and $b$. The above transformation is valid in the symmetric phase, but  can also be used in the broken phase, provided one first rotates the $z$-axis to bring it along the classical spin direction \cite{Dusuel05_2}.

%
%
%%%%%%%%%%%%%%%%%%%%%%%
%%%%%%%%%%%%%%%%%%%%%%%
\emph{The thermodynamic limit ---}
%%%%%%%%%%%%%%%%%%%%%%%
%%%%%%%%%%%%%%%%%%%%%%%
%
At fixed $\tau=L/N$, the Hamiltonian can be expanded in $1/N$. At order $(1/N)^0$ and for $h > 1$, one gets $H=NH^{(-1)}+H^{(0)}+\mathcal O(1/N)$ with $H^{(-1)}=-h/2$ and 
%
%
%%%%%%%%%%
\begin{eqnarray}
  \label{eq:H0}
  H^{(0)}&=&-\frac{1+\gamma}{4}+\frac{2h-\gamma-1}{2} \big( \cra \ana+\crb \anb \big)  \nonumber \\
&&+
\frac{\gamma-1}{4}  \Big[ \tau \big( {\cra}^2+\ana^2 \big) + (1-\tau) \big( {\crb}^2+\anb^2 \big) \nonumber \\ 
&&+
2 \sqrt{\tau(1-\tau)} \big( \cra \crb +\ana \anb \big) \Big].
\end{eqnarray}
%%%%%%%%%%
%
This effective bosonic Hamiltonian for the spin excitations is quadratic, and thus exactly solvable. The reduced density matrix can be written as
$\rho_{\mathcal{A}}={\rm e}^{-K}$  where, at the order we consider here, $K$ reads \cite{Peschel99,Peschel03}
%
%
%%%%%%%%%%
\begin{equation}
  \label{eq:K0}
  K^{(0)}=\kappa_0^{(0)}+\kappa_1^{(0)}\cra\ana+\kappa_2^{(0)}
  \Big( {\cra}^2+\ana^2 \Big)\,.
\end{equation}
%%%%%%%%%%
%
% 
The key ingredients leading to this form are: {\it i}) the eigenvalues of $\rho_\mathcal{A}$ are non-negative and smaller than one, which explains the exponential form; {\it ii}) Wick's theorem holds for quadratic Hamiltonians, constraining $K^{(0)}$ to be quadratic. The three coefficients $\kappa_i^{(0)}$ can be determined from the three conditions
%
%
%%%%%%%%%%
\begin{equation}
  \label{eq:constraints0}
  \mathrm{Tr}_\mathcal{A}\rho_\mathcal{A}=1,\;\;
  \langle\!\langle \cra\ana \rangle\!\rangle=\langle \cra\ana \rangle
  \mbox{ and }
  \langle\!\langle {\cra}^2 \rangle\!\rangle=\langle {\cra}^2 \rangle,
\end{equation}
%%%%%%%%%%
%
% 
where $\langle \Omega \rangle= \langle\psi|\Omega|\psi\rangle$ and
$\langle\!\langle \Omega \rangle\!\rangle =\mathrm{Tr}_\mathcal{A}({\rm e}^{-K}\Omega)$. To compute these expectation values, one simply has to diagonalize $H^{(0)}$ and $K^{(0)}$. Then, the $\kappa_i^{(0)}$'s are obtained by solving the $3\times3$ nonlinear system of equations (\ref{eq:constraints0}). From these coefficients, we finally obtain the entropy \cite{Barthel06_1}
%
%
%%%%%%%%%%
\begin{equation}
  \label{eq:entropy0}
  \mathcal{E}^{(0)}=\frac{\mu+1}{2}\ln\frac{\mu+1}{2} -\frac{\mu-1}{2}\ln\frac{\mu-1}{2},
\end{equation}
%%%%%%%%%%
%
% 
with
$\mu=\alpha^{-1/2}\sqrt{ \big[\tau\alpha+(1-\tau)\big] \big[\tau+\alpha(1-\tau)\big]}$,
and
%
%%%%%%%%%%%%%%%%%%%%
\begin{align}
\alpha&\textstyle=\sqrt{\frac{h-1}{h-\gamma}} \quad \mbox{for} \quad h>1, \\
\alpha&\textstyle=\sqrt{\frac{1-h^2}{1-\gamma}} \quad \mbox{for} \quad h<1. 
\end{align}
%%%%%%%%%%%%%%%%%%%%
%
It is interesting to note that $\alpha=\lim_{N\to\infty}4\langle S_y^2 \rangle/N$ also plays an important role for the concurrence which is a two-spin entanglement measure \cite{Dusuel04_3,Dusuel05_2}.
 
 In the broken phase $h<1$, the ground state is twofold degenerate in the thermodynamic limit. The two quantum states belong to the maximum spin sector $S=N/2$ and are eigenstates of the spin-flip operator $\prod_i \sigma_z^i$. This degeneracy is lifted for finite $N$.
Here, we calculated the entropy stemming from quantum fluctuations around one of the (fully polarized) classical ground states, which do not coincide with the quantum ground states just discussed. However, they are closely related and it turns out that the difference between $\mathcal{E}^{(0)}$ and the actual entropy is equal to $\ln 2$ \cite{Barthel06_3}.
At $h=\sqrt{\gamma}$, the entropy $\mathcal{E}^{(0)}$ vanishes since, there, the two degenerate quantum ground states can be chosen as sepa\-rable \cite{Dusuel05_2}.

\begin{figure}[t]
  \centering
  \includegraphics[width= 8cm]{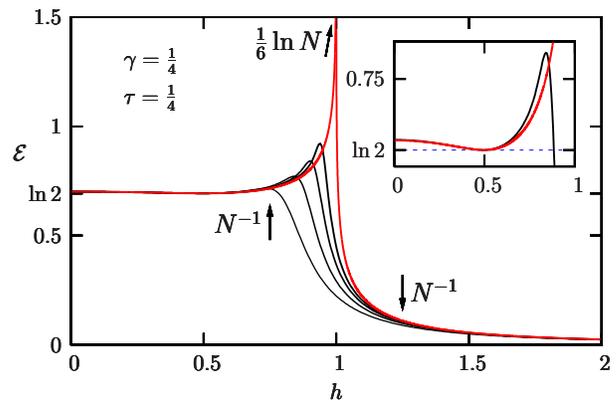}
  \caption{(Color online)
    Entanglement entropy as a function of the magnetic field, 
    with fixed $\gamma=1/4$, $\tau=1/4$ and $N= 32, 64, 128, 256$ (from numerics)
    and $\infty$ ($\mathcal{E}^{(0)}$). Arrows indicate the behavior of the 
    finite-size correction in various regions. For $h<1$, we have plotted $\mathcal{E}^{(0)}+\ln 2$ for the reason given in the text. The inset is a zoom around $h=\sqrt{\gamma}$ where $\mathcal{E}^{(0)}=0$, for $N=64$ (black line) and $\infty$ (red line).}
  \label{fig:entropy_vary_N}
\end{figure}

The comparison of $\mathcal{E}^{(0)}$ with numerical results obtained from exact diagonalization is shown in Fig.~\ref{fig:entropy_vary_N}. For large $N$, excellent agreement is observed in both phases.
%
%
%%%%%%%%%%%%%%%%
%%%%%%%%%%%%%%%%
%
%

%
%
%%%%%%%%%%%%%%%%%%%%%%%
%%%%%%%%%%%%%%%%%%%%%%%
\emph{Critical scaling of the entropy ---}
%%%%%%%%%%%%%%%%%%%%%%%
%%%%%%%%%%%%%%%%%%%%%%%
%
%
The main characteristic of the entropy $\mathcal{E}^{(0)}$ is that it is finite for $h \neq 1$ whereas it diverges at the critical point, in the vicinity of which one has 
%
%%%%%%%%%%%%%%%%%%%%
\begin{eqnarray}
\label{eq:entropyDL0}
\mathcal{E}^{(0)}&=&{-\frac{1}{4} \ln |h-1|+ \frac{1}{2}\ln [\tau(1-\tau)]+ \frac{1}{4}\ln (1-\gamma) } +\nonumber \\
&&1-x\ln 2+\mathcal O(|h-1|^{1/2}),
\end{eqnarray}
%%%%%%%%%%%%%%%%%%%%
%
with $x=1$ for $h>1$ and $x=5/4$ for $h<1$. The dependence on $h$ differs 
from the one given in Ref.~\cite{Latorre05_2} where, numerically, the entropy was found to behave as $-\frac{1}{6} \ln |h-1|$. The discrepancy comes from the too small investigated system sizes \cite{Barthel06_3}. 
It is also interesting to note that the same prefactor $-1/4$ was reported in the Dicke model \cite{Lambert04} using a different approach and it is very likely shared by a large class of collective models. 

As already observed for many physical quantities \cite{Dusuel04_3,Dusuel05_2}, the $1/N$ expansion of the entropy is singular at the critical point $h=1$. This is reminiscent of a nontrivial scaling behavior that we now discuss. 
We shall use the same scaling hypothesis as in Refs.~\cite{Dusuel04_3,Dusuel05_2} which assumes that in the vicinity of the critical point, a physi\-cal observable $\Phi$ can be written as the sum of a regular and a singular contribution, 
%
%%%%%%%%%%%%%%%%%%%%
\begin{equation}
  \Phi_N(h,\gamma)=
  \Phi_N^\mathrm{reg}(h,\gamma)+\Phi_N^\mathrm{sing}(h,\gamma)\,.
\end{equation}
%%%%%%%%%%%%%%%%%%%%
%
Here, singular means that  the function and/or its derivatives with respect to $h$ diverge at the critical point, following a power law. In addition, one has 
%
%%%%%%%%%%%%%%%%%%%%
\begin{equation}
\label{eq:phi_sing1}
  \Phi_N^\mathrm{sing}(h\simeq 1,\gamma) \sim   \frac{ (h-1)^{\xi_\Phi^h} (1-\gamma)^{\xi_\Phi^\gamma}} {N^{n_\Phi}}
  \mathcal{G}_\Phi (\zeta),
\end{equation}
%%%%%%%%%%%%%%%%%%%%
%
where the scaling variable $\zeta=N(h-1)^{3/2}(1-\gamma)^{-1/2}$ is introduced. 
The exponents $\xi_\Phi^h$, $\xi_\Phi^\gamma$ and $n_\Phi$ are characte\-ristics of the observables $\Phi$. 
Since no divergence can occur at finite $N$, one must have $\mathcal{G}_\Phi (\zeta) \sim \zeta^{-2 \xi_\Phi^h /3}$ and consequently
%
%%%%%%%%%%%%%%%%%%%%
\begin{equation}
\label{eq:phi_sing2}
\Phi_N^\mathrm{sing}(h=1,\gamma)\sim N^{-(n_\Phi + 2 \xi_\Phi^h /3)} (1-\gamma)^{\xi_\Phi^\gamma+ \xi_\Phi^h /3}.
\end{equation}
%%%%%%%%%%%%%%%%%%%%
%

To perform such an analysis for the entropy, one is led to consider $\Phi={\rm e}^\mathcal{E}$ which, contrary to $\mathcal{E}$, behaves as a power law at lowest order. Combining Eqs.~(\ref{eq:entropyDL0})-(\ref{eq:phi_sing2}), one readily identifies 
$\xi_\Phi^h=-\xi_\Phi^\gamma=-1/4$, and $n_\Phi=0$. In the large $N$ limit, one thus predicts
%
%%%%%%%%%%%%%%%%%%%%
\begin{equation}
\label{eq:entropy_critical}
\mathcal{E}(h=1) \sim \chi_N \ln N+ \chi_\gamma \ln (1-\gamma) + \chi_\tau \ln [\tau(1-\tau)],
\end{equation}
%%%%%%%%%%%%%%%%%%%%
%
with $\chi_N = \chi_\gamma=1/6$ and  $\chi_\tau =1/2$. This means that, at criticality and for fixed $N$, the entropy scales as $\frac{1}{2} \ln L$, just like in critical 1D systems. 
Although $\chi_\gamma$ is in agreement with previous numerical results \cite{Latorre05_2}, it is not the case for $\chi_\tau$ which was found to be close to $1/3$.

\begin{figure}[t]
  \centering
\includegraphics[width=8cm]{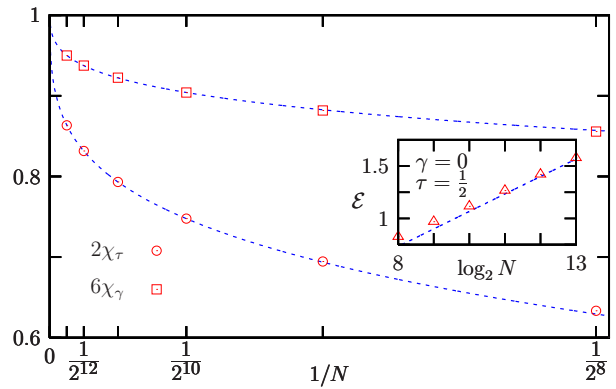}
  \caption{(Color online)
    Exponents $\chi_\gamma$ and $\chi_\tau$ as a function of $1/N$ obtained from numerical diagonalization of $H$. For clarity, we plotted $2 \chi_\tau$ and $6 \chi_\gamma$ which are expected to be equal to 1 in the thermodynamic limit (dotted lines are guides for the eyes). Inset: Entropy as a function of $\ln N$ at fixed $\gamma$ and $\tau$. The dotted line has a slope $\chi_N=1/6$.}
  \label{fig:exponents_N}
\end{figure}

To check our predictions, we have performed a finite-size scaling analysis of these prefactors.
As can be seen in Fig.~\ref{fig:exponents_N}, the exponents have not yet 
reached their asymptotic value at $N=2000$ which is the largest size analyzed 
in Ref.~\cite{Latorre05_2}. A simple extrapolation of these finite-size results to the thermodynamic limit confirms the predicted values of $\chi_N, \chi_\gamma$ and $\chi_\tau$ (dotted lines).

Our approach similarly predicts $\chi_N=1/6$ in the Dicke model which is consistent with the value of $0.14\pm 0.01$ obtained numerically \cite{Lambert04}.

%
%
%%%%%%%%%%%%%%%%
%%%%%%%%%%%%%%%%
%
%

\begin{figure}[t]
  \centering
\includegraphics[width=8cm]{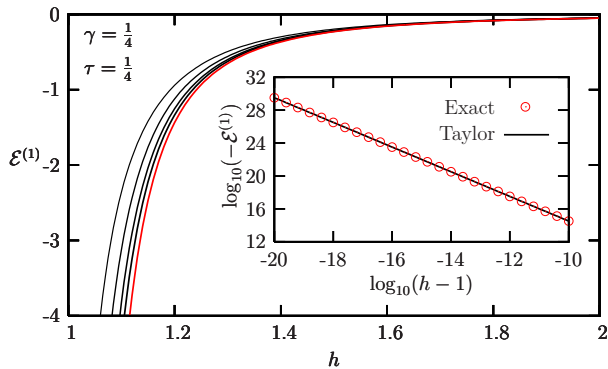}
  \caption{(Color online) Behavior of $N\big[\mathcal{E}_{\rm num}-\mathcal{E}^{(0)}\big]$ as a function of $h$ for fixed $\tau=1/4$ and $N=32,64,128,256$ (from numerics [black lines]) and $\infty$ ($\mathcal{E}^{(1)}$ obtained from the perturbative expansion [red lines]). Inset: Comparison between $\mathcal{E}^{(1)}$ (dots) and its Taylor expansion (\ref{eq:entropy_correction}) (solid line).
  }
  \label{fig:correction1}
\end{figure}

%
%
%%%%%%%%%%%%%%%%%%%%%%%
%%%%%%%%%%%%%%%%%%%%%%%
\emph{Finite-size corrections ---}
%%%%%%%%%%%%%%%%%%%%%%%
%%%%%%%%%%%%%%%%%%%%%%%
%
We shall now check the validity of the scaling hypothesis (\ref{eq:phi_sing1}).
To this purpose, we must at least compute the $1/N$ correction to the entropy, which requires to develop a perturbation theory to go beyond the free (quadratic) boson case. This calculation will be performed in the symmetric phase only, since this is sufficient to extract the scaling exponents. The approach we have developed in this aim constitutes the main contribution of the present work. First of all, one has to expand $H=N H^{(-1)}+H^{(0)}+H^{(1)}/N+\mathcal O(1/N^2)$, where
$H^{(1)}$ is quartic in $a$ and $b$.
The idea is then to expand $K=K^{(0)}+K^{(1)}/N+\mathcal O(1/N^2)$, with
%
%
%%%%%%%%%%
\begin{eqnarray}
  \label{eq:K1}
  K^{(1)}&=&
  \kappa_0^{(1)}+
  \kappa_1^{(1)}\cra\ana+
  \kappa_2^{(1)} \Big( {\cra}^2+\ana^2 \Big)+ 
  \kappa_3^{(1)}  {\cra}^2\ana^2+ \nonumber \\ &&
  \kappa_4^{(1)} \Big( {\cra}^3 \ana+\cra \ana^3 \Big)+
  \kappa_5^{(1)} \Big( {\cra}^4+\ana^4 \Big)\,.
\end{eqnarray}
%%%%%%%%%%
%
%
$K^{(1)}$ cannot contain any terms of order higher than four in $a$, because 
$H^{(0)}$ is quadratic and $H^{(1)}$ quartic in $a$ and $b$. In the framework of 
diagrammatic perturbation theory, all terms correspond to certain vertices. 
Tracing out mode $b$ cannot generate, in $K^{(1)}$, terms of sixth order in $a$ as 
every effective 6-legged vertex originates from the contraction of at least  
two bare 4-legged vertices and is thus of order $1/N^2$ or higher.
As in the quadratic case, the $\kappa_i^{(1)}$ can be determined from the conditions
%
%
%%%%%%%%%%
\begin{eqnarray}
  \label{eq:constraints1}
  &&\mathrm{Tr}_\mathcal{A}\rho_\mathcal{A}=1,\;\;
  \langle\!\langle \cra\ana \rangle\!\rangle=\langle \cra\ana \rangle,\;\;
  \langle\!\langle {\cra}^2 \rangle\!\rangle=\langle {\cra}^2 \rangle,\\
  &&\langle\!\langle {\cra}^2\ana^2 \rangle\!\rangle=\langle {\cra}^2\ana^2 \rangle,\;
  \langle\!\langle {\cra}^3 \ana \rangle\!\rangle=\langle {\cra}^3 \ana \rangle,\;
  \langle\!\langle {\cra}^4 \rangle\!\rangle=\langle {\cra}^4 \rangle, \nonumber
  \end{eqnarray}
%%%%%%%%%%
%
% 
which must be satisfied at order $1/N$. These expectation values can be evaluated perturbatively \cite{Barthel06_3}. An alternative route which we followed here is to compute them by {\it i}) diagonalizing the quartic operators $H^{(0)}+H^{(1)}/N$ and $K^{(0)}+K^{(1)}/N$ using the canonical transformation method described in Ref.~\cite{Vidal06_2} which requires to solve a system of 48 linear equations for $H$ and of 6 equations for $K$; {\it ii}) solving the resulting linear $6 \times 6$ system (\ref{eq:constraints1}) to obtain the coefficients $\kappa_i^{(1)}$. 
This second step can be done numerically, but the full exact solutions of this problem cannot be given explicitly. However, our main interest being the behavior of the entropy near the critical point, we have extracted its leading contribution which reads
%
%%%%%%%%%%%%%%%%%%%%
\begin{equation}
\label{eq:entropy_correction}
\mathcal{E}^{(1)}=-\frac{3 (1-\gamma)^{1/2}}{8(h-1)^{3/2}} +\mathcal O[(h-1)^{-1}].
\end{equation}
%%%%%%%%%%%%%%%%%%%%
%
This correction is in complete agreement with the scaling hypothesis (\ref{eq:phi_sing1}) since, at this order and in the vicinity of the critical point, one has
%
%
%%%%%%%%%%%%%%%%%%%%
\begin{eqnarray}
{\rm e}^\mathcal{E}&=& {\rm e}^{\mathcal{E}^{(0)}+\frac{1}{N}\mathcal{E}^{(1)}}\\
&\sim&[\tau(1-\tau)]^{1/2}\frac{(1-\gamma)^{1/4}}{(h-1)^{1/4}} \bigg[1-\frac{3 (1-\gamma)^{1/2}}{8 N (h-1)^{3/2}} \bigg]. \quad\nonumber
\end{eqnarray}
%%%%%%%%%%%%%%%%%%%%
%
%
As a check for the veracity of this perturbative method,  we compare in Fig.~\ref{fig:correction1} the numerical and analytical $1/N$ corrections to the entropy as functions of $h$. 
For increasing $N$, the numerical corrections converge quickly towards the analytical one. The inset is a check of the Taylor expansion (\ref{eq:entropy_correction}) in the vicinity of the critical point.
%
%
%%%%%%%%%%%%%%%%
%%%%%%%%%%%%%%%%
%
%

\emph{Discussion ---}
Away from criticality, the entanglement entropy of typical finite-dimensional 
systems is, on scales greater than the correlation length, proportional to 
the surface area of the considered subsystem \cite{Plenio05} (area law). 
At criticality, logarithmic corrections can occur \cite{Callan94,Wolf06,Gioev06,Barthel06_1}.
Non-critical collective models however behave differently, due to their 
infinite coordination number. 
In particular, the expansion of \eqref{eq:entropy0} for $h\neq 1$ and small $\tau=L/N$ yields the scaling $\mathcal{E}^{(0)}\propto - \tau \ln \tau + \mathcal O(\tau)$.
  The scaling  of the entropy (\ref{eq:entropy_critical}) in the critical 
collective model has to be compared to results obtained for other critical 
spin systems.  In 1D, the entropy at the critical point also scales 
logarithmically with the subsystem size $L$ with a prefactor depending on the 
universality class \cite{Latorre03}. For example, in the $XY$ model in a 
transverse field,  the entropy obeys a scaling law similar to  
(\ref{eq:entropy_critical}) but with differing prefactors 
\cite{Peschel04,Its05}. That the 1D scaling is also found in the critical 
collective model considered here, may seem surprising but is simply due to 
the fact that the ground state reduced density matrix is confined to the 
$S=L/2$ sector of the subsystem Hilbert space \cite{Latorre05_2}. 

Finally, let us emphasize that the perturbation theory we have developed and applied to the LMG model should not be associated to this model. It may be extended to more complex systems, provided one can determine $K^{(1)}$.

%
%
%%%%%%%%%%%%%%%%
\acknowledgments
%%%%%%%%%%%%%%%%
%
%
T.~B.\ thanks U.~Schollw\"ock for useful discussions, and the DFG for financial support.

%\bibliography{/Users/Julien/Documents/Tex/Bibliotheques/bibliotheque2}

\end{document}